\documentclass[twocolumn,showpacs,superscriptaddress]{revtex4}
\usepackage{amsmath}
\usepackage{amssymb}
\usepackage{bm}
\usepackage{epsfig}
\usepackage{graphicx}
\usepackage{color}
\usepackage[colorlinks=true]{hyperref} 
\usepackage[T2A]{fontenc}
\usepackage[cp1251]{inputenc}
\usepackage[russian,english]{babel}

\begin{document}

\newcount\timehh  \newcount\timemm
\timehh=\time \divide\timehh by 60
\timemm=\time
\count255=\timehh\multiply\count255 by -60 \advance\timemm by \count255

\title{
%Electron-spin-noise-detected nuclear spin dynamics\\
%Imaging nuclear spin dynamics by electron spin fluctuations\\
%Electron spin fluctuations reveal nuclear spin dynamics\\
%Nuclear spins made visible by electron spin noise spectroscopy\\
%Transient Overhauser resonances in electron spin noise spectra of n-type semiconductor nanostructures\\
%Single-shot pump-noise spectroscopy of n-GaAs microcavities reveals dynamics of the Overhauser field\\
Single-shot measurement of transient nuclear magnetization with spin-noise spectroscopy in n-GaAs microcavities }

\author{I.~I.~Ryzhov}
\author{S.~V.~Poltavtsev} 
\affiliation{Spin Optics Laboratory, St. Petersburg State University, 1 Ul'anovskaya,
Peterhof, St. Petersburg 198504, Russia}
\author{K.~V.~Kavokin}
\author{M.~M.~Glazov} 
\affiliation{Ioffe Institute, Russian Academy of Sciences, 26 Polytechnicheskaya,
St.-Petersburg 194021, Russia}
\affiliation{Spin Optics Laboratory, St. Petersburg State University, 1 Ul'anovskaya,
Peterhof, St. Petersburg 198504, Russia}
\author{G.~G.~Kozlov}
\affiliation{Spin Optics Laboratory, St. Petersburg State University, 1 Ul'anovskaya,
Peterhof, St. Petersburg 198504, Russia}
\author{M.~Vladimirova}
\author{D.~Scalbert}
\author{S.~Cronenberger}
\affiliation{Laboratoire Charles Coulomb UMR 5221 CNRS/Universit\'e de Montpellier,
Place Eugene Bataillon, 34095 Montpellier Cedex 05, France}
\author{A.~V.~Kavokin}
\affiliation{School of Physics and Astronomy, University of Southampton, SO17 1NJ
Southampton, United Kingdom}
 \affiliation{Spin Optics Laboratory, St. Petersburg State University, 1 Ul'anovskaya,
Peterhof, St. Petersburg 198504, Russia}
\author{A.~Lema\^{\i}tre}
\author{J. ~Bloch}
\affiliation{Laboratoire de Photonique et de Nanostructures, UPR
CNRS, Route de Nozay,
91460, Marcoussis, France}%
\author{V.~S.~Zapasskii}
\affiliation{Spin Optics Laboratory, St. Petersburg State University, 1 Ul'anovskaya,
Peterhof, St. Petersburg 198504, Russia}

\date{\today, file = \jobname.tex, printing time = \number\timehh\,:\,\ifnum\timemm<10 0\fi \number\timemm}

\begin{abstract}
We exploit spin noise spectroscopy (SNS) to directly observe build-up of dynamic nuclear polarization and relaxation of a perturbed nuclear spin-system to its equilibrium state in a single-shot experiment. The SNS experiments were performed on a layer of bulk $n$-type GaAs embedded into a high-finesse microcavity with negative detuning. The dynamic nuclear spin polarization is observed as a shift of the peak in the electron spin noise spectrum due to the build-up of the Overhauser field acting on the electron spin. The relaxation dynamics of nuclear spin system was revealed in the time-resolved SNS experiments where  the exponential decay of the Overhauser field  with characteristic  timescale of  hundreds of seconds was detected.   We show that elliptically polarized laser beam tuned in resonance with the cavity mode, whose energy corresponds to nominal transparency of the semiconductor, can nevertheless produce a sizable nuclear polarization. 
\end{abstract}
\maketitle

\emph{Introduction.} Host lattice nuclear spins take a special place among spin systems in semiconductors. On the one hand, exceptional robustness of nuclear spins to effects of environment opens up prospects to use nuclear spins for information processing~\cite{Boehme08062012}, while, on the other, they play major role in electron spin dynamics and decoherence~\cite{A.Greilich09282007,Foletti:2009aa,Bluhm:2011aa}. Weak interaction of nuclear spins with light strongly complicates direct optical studies of nuclear spins allowing, as a rule, indirect detection of nuclear spin state via spin dynamics of charge carriers, in which case the carriers provide feedback on nuclei, often making difficult the interpretation of experiments ~\cite{Urbaszek:2013ly}. An alternative method was proposed in~\cite{Giri2013}, where Faraday rotation induced by the Overhauser field was detected and used to trace the nuclear spin relaxation in a single-shot experiment. Though this method shows good promises for studying the nuclear spin dynamics, it gives relative values of nuclear fields, thus requiring complicated calibration procedures to determine the magnitude of nuclear polarization.  Here we demonstrate a method of visualization of nuclear spin dynamics based on monitoring of omnipresent electron spin fluctuations by means of the spin noise spectroscopy (SNS) technique, which has an advantage of giving absolute values of nuclear fields. SNS is a new method of research intended for studying magnetic resonance and spin dynamics of transparent paramagnets. Primarily demonstrated on atomic systems \cite{aleksandrov81,PhysRevLett.80.3487,Mitsui:2000nx,Crooker_Noise}, this technique is most widely applied, nowadays, to semiconductor systems, for which it proved to be highly efficient \cite{Oestreich:rev,Zapasskii:13}. The method is based on detection of the magnetization noise via fluctuations of the Faraday (Kerr) rotation of the system. Since the laser beam, in these measurements, usually probes the sample in the region of its transparency, this technique appears to be essentially nonperturbative, which was initially considered as its most important merit. However, an increase of polarimetric sensitivity of these measurements, which was achieved either by direct increase of the probe beam power \cite{Zapasskii:1982aa,Glasenapp:2013fk}, or by placing the sample inside a microcavity \cite{PhysRevB.89.081304}, the light power density on the sample substantially increased, and, at some point, effects of the light-induced perturbation became noticeable \cite{Glasenapp:2013fk,romer:103903,PhysRevB.79.035208,PhysRevB.83.155204,PhysRevA.83.032512}.

Already first experiments on electron SNS in semiconductor nanostructures have revealed nuclear spin effects~\cite{crooker2012}, which appear mainly as broadening of spin noise (SN) spectra caused by the frozen nuclear spin fluctuations experienced via hyperfine interaction by localized charge carriers~\cite{gi2012noise}. It was also predicted that nuclear spin polarization may result in drastic change of localized electrons SN~\cite{2014arXiv1412.0534S}. Here we propose to use Faraday-rotation-based SNS to monitor spin dynamics of optically pumped nuclei. The experiments are performed on the bulk $n$-type GaAs layer embedded into a high-finesse microcavity. The electron SN was detected via fluctuating Faraday rotation angle of a linearly polarized probe beam tuned to the cavity resonance about 20 meV below the bandgap of GaAs, i.e. nominally in the transparency region.  We demonstrate that the optical pumping of nuclear spins results in a shift of the electron SN peak, as if additional magnetic field $\bm B_N$, known as Overhauser field, was applied to the sample. We observe, in the electron SN spectra taken with time resolution of several seconds, the build-up of the nuclear spin polarization under pumping as well as relaxation of nuclear spins towards disordered state after pump has been switched off. Surprisingly, we could pump nuclear spins not only by interband optical excitation, but also by the low-wavelength probe beam itself, when it was made elliptically polarized and increased in intensity. Measuring transient nuclear fields in GaAs samples with donor concentrations below ($n_D=2*10^{15}$) and above ($n_D=4*10^{16}$) the insulator-to-metal transition have revealed different mechanisms of dynamic polarization and relaxation of nuclear spins in semiconductors with localized and delocalized (Fermi-edge) electrons.

\begin{figure}
 \includegraphics[width=\columnwidth,clip]{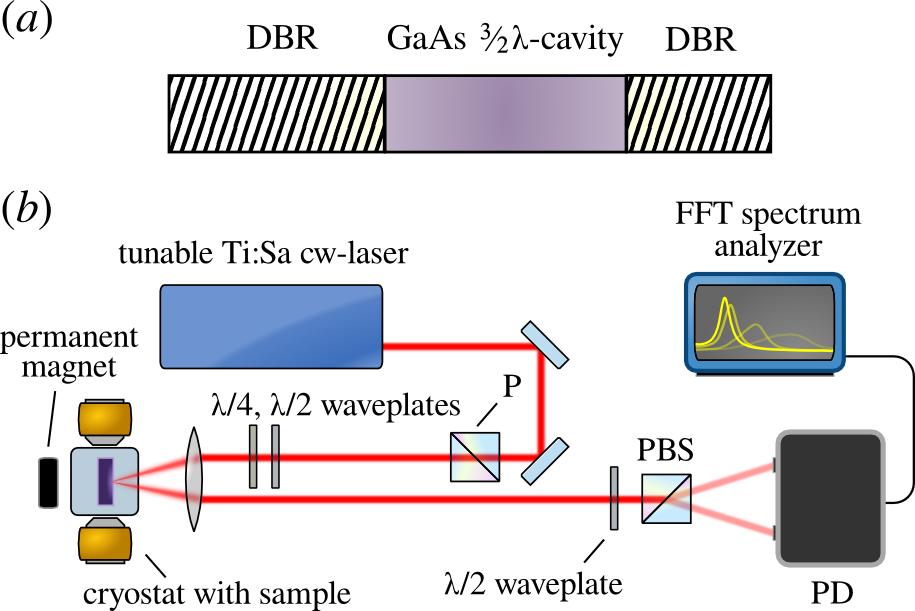}
 \caption{Schematic representation of the studied sample (a) and setup (b). P --- polarizer, PBS --- polarizing beamsplitter, PD --- balanced photodetector.}
  \label{fig:ss}
  \end{figure}

\emph{Sample and experimental setup.} The schemes of the sample and the setup are shown in Fig.~\ref{fig:ss}, panels (a) and (b), respectively. The experiments were performed with a layer of bulk $n$-type GaAs embedded in a  $3\lambda/2$ microcavity with a high-finesse, quality factor $Q \sim 10^4$ similar to the ones used in Refs.~\cite{PhysRevLett.111.087603, PhysRevB.85.195313}. The electron density in the sample $n\approx 4\times 10^{16}$~cm$^{-3}$ was chosen to realize longest possible spin relaxation times~\cite{Kikkawa98,Dzhioev02} and, correspondingly, narrowest electron SN spectrum widths. The sample was mounted in the closed-cycle cryostat ``Montana'' providing temperatures down to $3$~K and transverse (with respect to the optical axis) magnetic fields (Voight geometry) up to $0.7$~T. Additional longitudinal magnetic field (of about several mT) could be applied to the sample using a permanent magnet. As the probe and pump beam we used one and the same output emission of a cw Ti-Sapphire laser tuned to the cavity mode of the structure ($\lambda\approx 833$~nm), corresponding to the region of nominal transparency of the sample. Spectral tuning was performed by choosing a point on the gradient sample.  The size of the spot of the focused beam on the sample was $\sim $ 30 $\mu$m and its power varied from $0.25$~mW up to $6$~mW. Ellipticity of the beam was controlled using the quarter-wave plate (Fig. 1,a). The SN spectra were measured by means of a balanced photodetector with the bandwidth 200~MHz and a FFT spectrum analyzer. The accumulation time of the system was usually a few seconds and, therefore, we were able to study time evolution of the SN spectra making measurements each several seconds.

\begin{figure}
 \includegraphics[width=.8\columnwidth,clip]{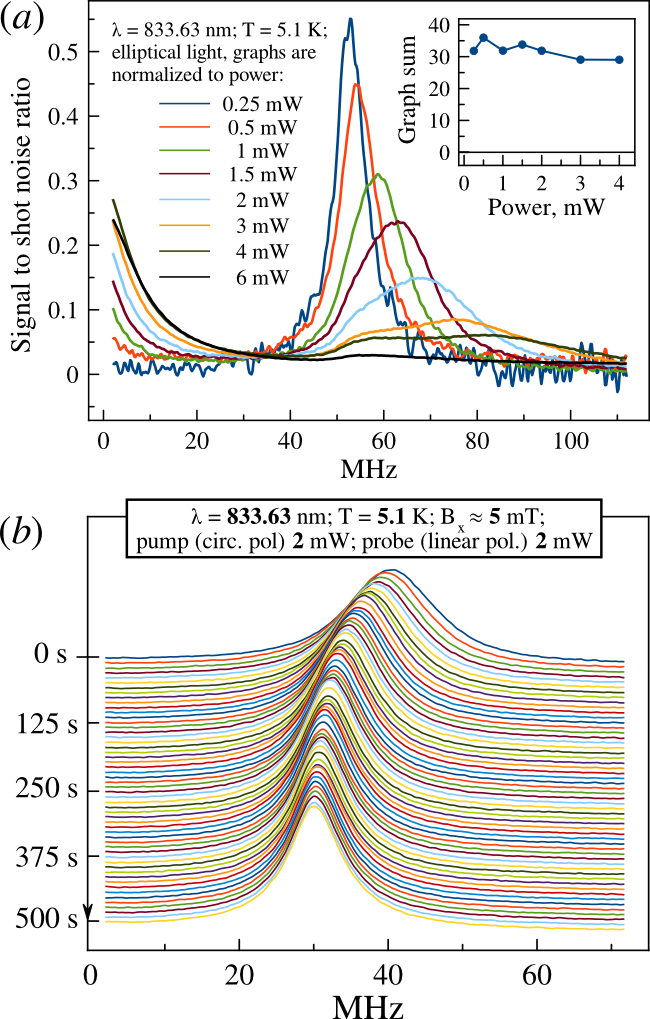}
 \caption{(a) Kerr rotation noise spectra measured at $B_x=8.5$~mT, $T=5.1$~K by elliptically polarized probe with $P_{circ} =20$\%. The spectra are normalized to the probe power. An inset shows the integral of the spectra as a function of probe power. (b) Kerr rotation noise spectra measured at $B_x=4$~mT, $T=5.1$~K each 7 seconds (curves from top to bottom, offset for clarity) after illumination of the sample by circularly polarized light by 5 minutes.}
  \label{fig:ellipt}
  \end{figure}

\emph{Experimental results.} Figure~\ref{fig:ellipt}(a) shows the SN spectra measured at a transverse magnetic field $B_x=...$~mT at low temperature of about 3~K by slightly elliptical probe beam ($P_{circ} = (I_+-I_-)/(I_++I_-)=0.2$) for different beam powers. Polarization plane of the probe beam was aligned along the axis of residual birefringence of the sample (which could be checked by absence of any nonlinearities related to admixture of circular polarization). At small probe power (0.25~mW), we observed a narrow isolated line at a frequency corresponding to the $g$-factor  $|g| = 0.435$, which well correlates with the known value of $g$-factor for the bulk GaAs and can be attributed to the spontaneous electron spin resonance detected by the SNS. The line has a Lorentzian shape with the half width at half maximum of $\delta \nu \approx 2$~MHz corresponding to the electron spin relaxation time $\tau_s = 75$~ns, and consistent with previous studies~\cite{Kikkawa98,Dzhioev02}  

An increase of the probe power results in the increase of the measured noise proportionally to the probe power, so that the area of the spectra normalized by the probe power remains approximately constant (see inset in Fig.~\ref{fig:ellipt}(a)), demonstrating the absence of noticeable generation of electrons whose noise we detect. The SN line exhibits strong broadening (which can be ascribed to inhomogeneity of pumping over the light spot) and a pronounced shift towards higher frequencies with increasing probe intensity. In addition, a peak at zero-frequency, $\nu=0$, arises, whose magnitude increases approximately in proportion with the probe power and whose width is close to that of the Lorentzian peak at low powers. Changing sign of $P_{circ}$ does not change the SN spectrum.

To have a deeper insight into the observed nonlinearitues, we have studied the effect of {\it circularly polarized} probe beam on the SN spectrum of the sample. In this case, the circularly polarized ($P_{circ} = 1$) probe beam of about 2 mW in power illuminated the sample for several minutes. After that, the quarter-wave plate, Fig~\ref{fig:ss}(b), was removed, and the SN spectra were recorded, in a conventional way, each 5 seconds. The set of the spectra are presented in Fig.~\ref{fig:ellipt}(b), the curves are offset along vertical axis for clarity. We observe that the peak at $\nu=0$ is absent, while the position of the spin-resonance line initially appears to be strongly shifted towards higher frequencies. Its position depends, in particular, on the probe power and illumination time as well as on the magnetic field. However, with time this line shifts towards the non-perturbed position defined by the electron $g$-factor and external magnetic field.

\emph{Discussion.} The experimental results presented in Fig.~\ref{fig:ellipt} demonstrate that under illumination of our sample by elliptical and, particularly, circularly polarized radiation the effective magnetic fields acting on electron spin builds-up. Particularly, an effective field along the external field $B_x$ appears, which, in the absence of illumination, decays with the macroscopic time constant of hundreds of seconds. The appearance of such a field evidently corresponds to the polarization of some spin system. Since electron spin relaxation time amounts to just several nanoseconds, the only candidate for the spins responsible for the effect is the system of host lattice nuclear spins, which, for similar samples, can demonstrate the (longitudinal, i.e. for the component parallel to the external field) spin relaxation times $T_{1,N}\sim 10^2$~sec, see e.g.~\cite{PhysRevLett.111.087603}. 

To demonstrate how the non-equilibrium nuclear polarization builds-up in our experiment, we  assume that, due to residual absorption in the sample, the electrons are spin-polarized, and the average electron spin $\bm S \ne 0$. In accordance with the general theory~\cite{abragam,opt_or_book}, we assume that the system of the host lattice nuclei is well isolated from the electronic system ($T_{1,N} \gg T_{2,N} \sim 10^{-4}$~sec, where $T_{2,N}$ is the nuclear transverse relaxation time~\cite{opt_or_book,dyakonov_book}) and introduce the nuclear spin temperature $\Theta$, which can be found from the energy balance~\cite{opt_or_book}:
\begin{equation}
\label{balance}
\frac{1}{\Theta} = \frac{4I}{\mu_I}\frac{(\bm S \cdot \bm B) + b_e S^2}{(\bm B + b_e \bm S)^2 + \xi B_L^2}.
\end{equation}
Here $I$ is the nuclear magnetic moment ($I=3/2$ both for Ga and As isotopes), $\mu_I$ is the nuclear magneton, $\bm B ||x$ is the external magnetic field acting on nuclear spins, $b_e$ describes the strength of the hyperfine interaction and the Knight field acting on the nuclear spins is $b_e \bm S$, $B_L$ is the characteristic value of the local field and $\xi$ is the constant determined by the spin-spin interactions. Equation~\eqref{balance} takes into account that the nuclear spin system is affected by the total field $\bm B + b_e \bm S$ being the sum of both external magnetic field $\bm B$ and Knight field $b_e \bm S$. We recall that in the framework of the nuclear spin temperature the average nuclear spin $\bm I = (I+1) \mu_I (\bm B + b_e \bm S)/3\Theta$ which results in the formation of the Overhauser field acting on electron spins 
\begin{equation}
\label{overhauser}
\bm B_N = b_N \frac{\bm I}{I}.
\end{equation}
Equation~\eqref{balance} demonstrates that the nuclear spin cooling and dynamical spin polarization arises even at $\bm S \perp \bm B$ due to the effect of Knight field. In this case sign of the nuclear spin temperature $\Theta$ is independent of the sign of the circular polarization. In agreement with general theory~\cite{opt_or_book} for GaAs the Overhauser field $\bm B_N$ has the same direction as the external field resulting in an increase of the electron spin precession frequency, see Fig.~\ref{fig:ellipt}. Taking into account that the detected value of the Overhauser field is about $10^{-3}$ of the maximum value we obtain the nuclear spin temperature $|\Theta|\sim 1$~mK.

\begin{figure}
\includegraphics[width=.9\columnwidth,clip]{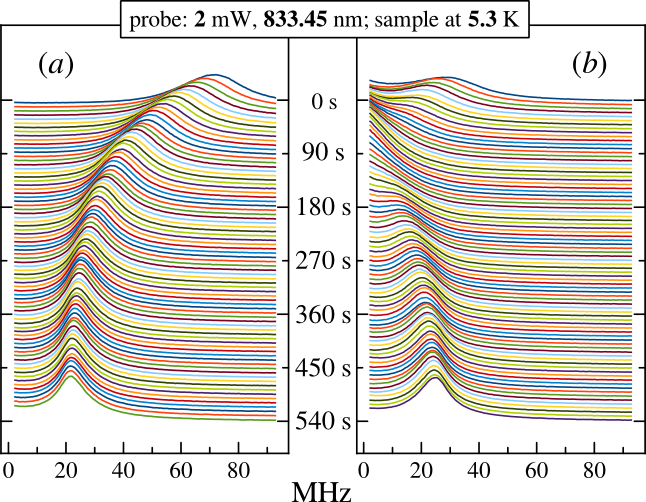}
 \caption{Kerr rotation noise spectra measured at $B_x=4$~mT, $T=5.3$~K each 7 seconds (curves from top to bottom, offset for clarity) after illumination of the sample by $\sigma^+$ circularly polarized light by 5 minutes in the presence of longitudinal magnetic field $B_z \approx +12$~mT (a) and $B_z \approx -12$~mT (b). }
  \label{fig:Bz}
  \end{figure}

In order to control the direction of nuclear spin $\bm I$ and, hence, the Overhauser field $\bm B_N$, we applied an additional magnetic field $B_z\sim 10$ ~mT along $z$-axis. Provided that it exceeds by far the Knight field (the latter is on the order of $5\times 10^{-2}$~mT~\cite{PhysRevB.85.195313}), $|B_z|>|b_e S|$, the sign of the nuclear spin temperature is, in this case, given by the product $B_z S_z\propto P_{circ}$ and can be reversed by changing the sign of the radiation helicity or the sign of $B_z$ for the fixed helicity. Longitudinal magnetic field was removed after the pumping stage, and nuclear spins were rapidly (during the time $\sim T_{2,N}$) oriented parallel or anitparallel to $B_z$ depending on the sign of $\Theta$. The results of the experiments carried out for $B_z>0$ and $B_z<0$ are shown in Fig.~\ref{fig:Bz}, panels (a) and (b), respectively. The results presented in Fig.~\ref{fig:Bz}(a) are qualitatively similar to those in Fig.~\ref{fig:ellipt}(b): In agreement with Eq.~\eqref{balance} in both cases the Overhauser field adds up with the external field. One can see that in the presence of external longitudinal magnetic field the dynamical nuclear polarization builds-up much more efficiently compared with the field applied in the Voight geometry: The shift of the SN peak exceeds 30~MHz in the former case as compared with less than 10~MHz in the latter.

The case of $B_z<0$, Fig.~\ref{fig:Bz}(b) is qualitatively different, showing a nontrivial dynamics of the electron SN peak. First it goes from about 40 MHz to 0 and then restores back to its unperturbed frequency of $\approx 20$~MHz. This result can be easily understood taking into account that the SN spectrum is symmetric with respect to $\nu\to -\nu$~\cite{gi2012noise}, therefore non-monotonic shift of the SN peak demonstrates that the total field acting on the electron spin $B + B_N$ changes it sign. Indeed, in agreement with Eq.~\eqref{balance} for $P_{circ}>0$ and $B_z<0$ the $\bm B_N$ is antiparallel to $\bm B$ and, for sufficiently low nuclear spin temperature, the overcompensation may occur~\cite{PhysRevB.85.195313}. In the process of nuclear spin relaxation, we come to the point $B_N=-B$ when the electron spin resonance line visualized by the SNS passes through zero frequency.

Finally, we address possible origins of the peak at $\nu=0$ in the SN spectra observed in the presence of elliptically polarized probe. This peak immediately bulds up if the beam ellipticity appears and it disappears if $P_{circ}$ is put to zero, its magnitude is independent of the illumination time. Therefore, nuclear spin effects as possible origin of the zero frequency peak can be excluded with high probability. Hence, the peak can be related with an effect of elliptically polarized light on the electron ensemble. The appearance of zero-frequency peak can be tentatively assigned to the appearance of the effective optically-induced field acting on electron spins. Taking into account the fact that the peak amplitude scales linearly with radiation power $P$ and making use of the cubic symmetry of GaAs we can write for this field $\bm B_{opt} = \varkappa P_{circ} P \bm e_z$, where $\bm e_z$ is the unit vector along the light propagation axis $z$ and $\varkappa$ is the coefficient. In the presence of $\bm B_{opt}$ electron spins precess in the total field $\bm B + \bm B_{opt}$ (Overhauser field is neglected for simplicity), which now has non-zero $z$-component. The electron spin fluctuations precess now in the tilted field and the zero-frequency component in the electron spin noise spectrum appears~\cite{gi2012noise} in the agreement with the experiment in Fig.~\ref{fig:ellipt}(a). One of the evident candidates for $\bm B_{opt}$ is the ac Zeeman effect induced by the circular field in analogy to ac Stark effect~\cite{Chemla1989233}: Off-resonant circularly polarized field yields the effective Zeeman splitting of electron levels in the form~\cite{Chemla1989233,PhysRev.143.574}: $g\mu_B B_{opt} = 2(|d_+|^2 - |d_-|^2) P_{circ} |E_0|^2/|\delta|$, where $d_\pm$ are the dipole moments for the transition to the $s_z=+1/2$ electron spin state in $\sigma^+$ and $\sigma^-$ polarizations, respectively, $E_0$ is the amplitude of electromagnetic field, and $\delta$ is the detuning from the resonance. Our estimates show that for experimental parameters and the nominal cavity $Q$-factor of $10^4$ the $B_{opt}$ for free electrons can reach values of $\sim 1$~T, which exceeds by far the value estimated from Fig.~\ref{fig:ellipt}(a) of several mT. The origin of this discrepancy is not clear, it is probably related with (i) the fact that the SN signal comes from localized electrons on a donor pairs~\cite{PhysRevLett.111.087603} whose optical dipole moment is smaller than for transitions to free electron states and (ii) the fact that real $Q$-factor of the cavity is smaller than the nominal one~\footnote{Note that inverse Faraday effect is excluded by the measurement scheme because only fluctuating part of Kerr signal is detected.}. 

\emph{Conclusion.} 
In this paper, we demonstrate a new aspect of SNS that implies combination of linear and nonlinear spin noise spectroscopy to directly observe nuclear spin dynamics in real time. By using the circularly polarized beam to polarize host lattice nuclei in bulk GaAs and the same but linearly polarized beam to monitor electron spin fluctuations with time resolution we are able to detect nuclear spin dynamics and, particularly, longitudinal nuclear relaxation time. The suggested SNS-based method of the nuclear spin dynamics visualization may pave a way for studying the nuclear spin dynamics without noticeable perturbation of both nuclear and electron spin systems.

%\cite{Chekhovich:2013ys}

%\cite{Peddibhotla:2013aa}

%\cite{PhysRevLett.111.087603}

%\bibliographystyle{misha}
%\bibliography{/Users/misha/Work/Coherent/Bibliography/all-1}

\begin{thebibliography}{10}
\providecommand{\selectlanguage}[1]{\relax}
%\input{babelbst.tex}
\newcommand{\Capitalize}[1]{\uppercase{#1}}
\newcommand{\capitalize}[1]{\expandafter\Capitalize#1}

\bibitem{Boehme08062012}
C.~Boehme, D.~R. McCamey.
\newblock \emph{Nuclear-Spin Quantum Memory Poised to Take the Lead}.
\newblock Science \textbf{336}, 1239 (2012).

\bibitem{A.Greilich09282007}
A.~Greilich, A.~Shabaev, D.~R. Yakovlev, A.~L. Efros, I.~A. Yugova, D.~Reuter,
  A.~D. Wieck, M.~Bayer.
\newblock \emph{{Nuclei-induced frequency focusing of electron spin
  coherence}}.
\newblock Science \textbf{317}, 1896 (2007).

\bibitem{Foletti:2009aa}
S.~Foletti, H.~Bluhm, D.~Mahalu, V.~Umansky, A.~Yacoby.
\newblock \emph{Universal quantum control of two-electron spin quantum bits
  using dynamic nuclear polarization}.
\newblock Nat Phys \textbf{5}, 903 (2009).

\bibitem{Bluhm:2011aa}
H.~Bluhm, S.~Foletti, I.~Neder, M.~Rudner, D.~Mahalu, V.~Umansky, A.~Yacoby.
\newblock \emph{Dephasing time of GaAs electron-spin qubits coupled to a
  nuclear bath exceeding 200$\mu$s}.
\newblock Nat. Phys. \textbf{7}, 109 (2011).

\bibitem{Urbaszek:2013ly}
B.~Urbaszek, X.~Marie, T.~Amand, O.~Krebs, P.~Voisin, P.~Maletinsky,
  A.~H\"ogele, A.~Imamoglu.
\newblock \emph{Nuclear spin physics in quantum dots: An optical
  investigation}.
\newblock Rev. Mod. Phys. \textbf{85}, 79 (2013).


\bibitem{Giri2013}
R. ~Giri, S.~Cronenberger, M. ~M. ~Glazov, K. ~V.~ Kavokin, A. ~Lema�tre, J. ~Bloch, M.~ Vladimirova, and D. ~Scalbert.
\newblock \emph{Nondestructive Measurement of Nuclear Magnetization by Off-Resonant Faraday Rotation}.
\newblock Phys. Rev. Lett \textbf{111}, 087603 (2014).

\bibitem{aleksandrov81}
E.~Aleksandrov, V.~Zapasskii.
\newblock \emph{Magnetic resonance in the {F}araday-rotation noise spectrum}.
\newblock JETP \textbf{54}, 64 (1981).

\bibitem{PhysRevLett.80.3487}
J.~L. S\o{}rensen, J.~Hald, E.~S. Polzik.
\newblock \emph{Quantum Noise of an Atomic Spin Polarization Measurement}.
\newblock Phys. Rev. Lett. \textbf{80}, 3487 (1998).

\bibitem{Mitsui:2000nx}
T.~Mitsui.
\newblock \emph{Spontaneous Noise Spectroscopy of an Atomic Magnetic
  Resonance}.
\newblock Phys. Rev. Lett. \textbf{84}, 5292 (2000).

\bibitem{Crooker_Noise}
S.~A. Crooker, D.~G. Rickel, A.~V. Balatsky, D.~L. Smith.
\newblock \emph{Spectroscopy of spontaneous spin noise as a probe of spin
  dynamics and magnetic resonance}.
\newblock Nature \textbf{431}, 49 (2004).

\bibitem{Oestreich:rev}
J.~H{\"u}bner, F.~Berski, R.~Dahbashi, M.~Oestreich.
\newblock \emph{The rise of spin noise spectroscopy in semiconductors: From
  acoustic to GHz frequencies}.
\newblock physica status solidi (b) \textbf{251}, 1824 (2014).

\bibitem{Zapasskii:13}
V.~S. Zapasskii.
\newblock \emph{Spin-noise spectroscopy: from proof of principle to
  applications}.
\newblock Adv. Opt. Photon. \textbf{5}, 131 (2013).

\bibitem{Zapasskii:1982aa}
V.~Zapasskii.
\newblock \emph{Highly sensitive polarimetric
  techniques (review)}.
\newblock Journal of Applied Spectroscopy \textbf{37}, 857 (1982).

\bibitem{Glasenapp:2013fk}
P.~Glasenapp, A.~Greilich, I.~I. Ryzhov, V.~S. Zapasskii, D.~R. Yakovlev, G.~G.
  Kozlov, M.~Bayer.
\newblock \emph{Resources of polarimetric sensitivity in spin noise
  spectroscopy}.
\newblock Phys. Rev. B \textbf{88}, 165314 (2013).

\bibitem{PhysRevB.89.081304}
S.~V. Poltavtsev, I.~I. Ryzhov, M.~M. Glazov, G.~G. Kozlov, V.~S. Zapasskii,
  A.~V. Kavokin, P.~G. Lagoudakis, D.~S. Smirnov, E.~L. Ivchenko.
\newblock \emph{Spin noise spectroscopy of a single quantum well microcavity}.
\newblock Phys. Rev. B \textbf{89}, 081304 (2014).

\bibitem{romer:103903}
M.~Romer, J.~Hubner, M.~Oestreich.
\newblock \emph{Spin noise spectroscopy in semiconductors}.
\newblock Review of Scientific Instruments \textbf{78}, 103903 (2007).

\bibitem{PhysRevB.79.035208}
S.~A. Crooker, L.~Cheng, D.~L. Smith.
\newblock \emph{Spin noise of conduction electrons in $n$ -type bulk {G}a{A}s}.
\newblock Phys. Rev. B \textbf{79}, 035208 (2009).

\bibitem{PhysRevB.83.155204}
Q.~Huang, D.~S. Steel.
\newblock \emph{Optical excitation effects on spin-noise spectroscopy in
  semiconductors}.
\newblock Phys. Rev. B \textbf{83}, 155204 (2011).

\bibitem{PhysRevA.83.032512}
W.~Chalupczak, R.~M. Godun.
\newblock \emph{Near-resonance spin-noise spectroscopy}.
\newblock Phys. Rev. A \textbf{83}, 032512 (2011).

\bibitem{crooker2012}
Y.~Li, N.~Sinitsyn, D.~L. Smith, D.~Reuter, A.~D. Wieck, D.~R. Yakovlev,
  M.~Bayer, S.~A. Crooker.
\newblock \emph{Intrinsic Spin Fluctuations Reveal the Dynamical Response
  Function of Holes Coupled to Nuclear Spin Baths in (In,Ga)As Quantum Dots}.
\newblock Phys. Rev. Lett. \textbf{108}, 186603 (2012).

\bibitem{gi2012noise}
M.~M. Glazov, E.~L. Ivchenko.
\newblock \emph{Spin noise in quantum dot ensembles}.
\newblock Phys. Rev. B \textbf{86}, 115308 (2012).

\bibitem{2014arXiv1412.0534S}
D.~S. {Smirnov}.
\newblock \emph{{Spin noise of localized electrons interacting with optically
  cooled nuclei}}.
\newblock ArXiv e-prints  (2014).

\bibitem{PhysRevLett.111.087603}
R.~Giri, S.~Cronenberger, M.~M. Glazov, K.~V. Kavokin, A.~Lema\^{i}tre,
  J.~Bloch, M.~Vladimirova, D.~Scalbert.
\newblock \emph{Nondestructive Measurement of Nuclear Magnetization by
  Off-Resonant Faraday Rotation}.
\newblock Phys. Rev. Lett. \textbf{111}, 087603 (2013).

\bibitem{PhysRevB.85.195313}
R.~Giri, S.~Cronenberger, M.~Vladimirova, D.~Scalbert, K.~V. Kavokin, M.~M.
  Glazov, M.~Nawrocki, A.~Lema\^{i}tre, J.~Bloch.
\newblock \emph{Giant photoinduced Faraday rotation due to the spin-polarized
  electron gas in an $n$-GaAs microcavity}.
\newblock Phys. Rev. B \textbf{85}, 195313 (2012).

\bibitem{Kikkawa98}
J.~M. Kikkawa, D.~D. Awschalom.
\newblock \emph{Resonant Spin Amplification in $n$-Type $\mbox{GaAs}$}.
\newblock Phys. Rev. Lett. \textbf{80}, 4313 (1998).

\bibitem{Dzhioev02}
R.~I. Dzhioev, K.~Kavokin, V.~Korenev, M.~Lazarev, B.~Y. Meltser, M.~N.
  Stepanova, B.~P. Zakharchenya, D.~Gammon, D.~S. Katzer.
\newblock \emph{Low-temperature spin relaxation in $n$-type $\mbox{GaAs}$}.
\newblock Phys. Rev. B \textbf{66}, 245204 (2002).

\bibitem{abragam}
A.~Abragam.
\newblock {\selectlanguage{english}\emph{Principles of Nuclear Magnetism}}
  (Oxford Science Publications, 2002).

\bibitem{opt_or_book}
F.~Meier, B.~Zakharchenya (eds).
\newblock \emph{Optical orientation} (Horth-Holland, Amsterdam, 1984).

\bibitem{dyakonov_book}
M.~I. Dyakonov (ed.).
\newblock \emph{Spin physics in semiconductors} (Springer-Verlag: Berlin,
  Heidelberg, 2008).

\bibitem{Chemla1989233}
D.~Chemla, W.~Knox, D.~Miller, S.~Schmitt-Rink, J.~Stark, R.~Zimmermann.
\newblock \emph{The excitonic optical stark effect in semiconductor quantum
  wells probed with femtosecond optical pulses}.
\newblock Journal of Luminescence \textbf{44}, 233  (1989).

\bibitem{PhysRev.143.574}
P.~S. Pershan, J.~P. van~der Ziel, L.~D. Malmstrom.
\newblock \emph{Theoretical Discussion of the Inverse $\mbox{F}$araday Effect,
  $\mbox{R}$aman Scattering, and Related Phenomena}.
\newblock Phys. Rev. \textbf{143}, 574 (1966).

\bibitem{Chekhovich:2013ys}
E.~A. Chekhovich, M.~M. Glazov, A.~B. Krysa, M.~Hopkinson, P.~Senellart,
  A.~Lemaitre, M.~S. Skolnick, A.~I. Tartakovskii.
\newblock \emph{Element-sensitive measurement of the hole-nuclear spin
  interaction in quantum dots}.
\newblock Nat Phys \textbf{9}, 74 (2013).

\bibitem{Peddibhotla:2013aa}
P.~Peddibhotla, F.~Xue, H.~I.~T. Hauge, S.~Assali, E.~P. A.~M. Bakkers,
  M.~Poggio.
\newblock \emph{Harnessing nuclear spin polarization fluctuations in a
  semiconductor nanowire}.
\newblock Nat Phys \textbf{9}, 631 (2013).

\end{thebibliography}

\end{document}